\documentclass[doublecol]{epl2}

\begin{document}

\title{Infinite-noise criticality: Nonequilibrium phase transitions in fluctuating environments}

\author{Thomas Vojta\inst{1} \and Jos\'e A. Hoyos\inst{2}}

\institute{
  \inst{1} Department of Physics, Missouri University of Science and Technology, Rolla, MO 65409, USA\\
  \inst{2} Instituto de F\'{i}sica de S\~ao Carlos, Universidade de S\~ao Paulo,
           C.P. 369, S\~ao Carlos, S\~ao Paulo 13560-970, Brazil
}

\abstract{
We study the effects of time-varying environmental noise on nonequilibrium phase transitions in
spreading and growth processes. Using the examples of the logistic evolution equation as well as
the contact process, we show that such temporal disorder gives rise to a distinct type of critical points
at which the effective noise amplitude diverges on long time scales. This leads to enormous density fluctuations
characterized by an infinitely broad probability distribution at criticality. We develop a real-time
renormalization-group theory that provides a general framework for the effects of temporal disorder on
nonequilibrium processes. We also discuss how general this exotic critical behavior is,  we illustrate
the results by computer simulations, and we touch upon experimental applications of our theory.
}

\maketitle

\section{Introduction}

Systems far from thermodynamic equilibrium can undergo
abrupt transitions between different nonequilibrium steady states. These nonequilibrium
phase transitions feature collective behavior over long distances and
times just as thermodynamic equilibrium transitions. Examples are found
in the extinction dynamics of epidemics and bio-populations as well as in
surface growth, turbulent flow, and catalytic reactions
\cite{SchmittmannZia95,MarroDickman99,Hinrichsen00,Odor04,Luebeck04,
TauberHowardVollmayrLee05,HenkelHinrichsenLuebeck_book08}.

Nonequilibrium processes often occur in spatially inhomogeneous systems and
time-varying environments, i.e.,  in the presence of spatial and temporal disorder.
Spatial disorder can have dramatic effects on nonequilibrium
transitions.
For example, it destroys the ubiquitous directed percolation (DP) \cite{GrassbergerdelaTorre79} universality class and produces an exotic
infinite-randomness critical point
\cite{HooyberghsIgloiVanderzande03,HooyberghsIgloiVanderzande04,VojtaDickison05,VojtaFarquharMast09,Vojta12,OliveiraFerreira08,Hoyos08}.
The associated Griffiths phases are dominated by rare fluctuations
and feature anomalous
power-law relaxation \cite{Noest86,Noest88}. Similar behavior is found
in percolating \cite{VojtaLee06,LeeVojta09} and quasiperiodic
\cite{BarghathiNozadzeVojta14} systems (for a review, see Ref.\ \cite{Vojta06}).

The effects of environmental noise, i.e., temporal disorder, have attracted
less attention. Kinzel \cite{Kinzel85} showed that temporal disorder destabilizes the DP transition
because its correlation time exponent $\nu_\parallel=\nu_\perp z$ violates the generalized
Harris criterion $\nu_\parallel \ge 2$.  Jensen \cite{Jensen96,Jensen05}
employed series expansions and Monte-Carlo simulations to determine the fate of the
DP transition with temporal disorder and reported nonuniversal critical exponents.
In an intriguing paper, Vazquez et al.\ \cite{VBLM11}
demonstrated that rare noise fluctuations can lead to a temporal
analog of the Griffiths phase, featuring an unusual power-law relation between lifetime
and system size. Environmental noise
has also been studied within space-independent (single-variable) models of biological population
dynamics (see, e.g., Refs.\ \cite{Leigh81,KamenevMeersonShklovskii08,OvaskainenMeerson10}).
Despite these efforts, a general framework for understanding nonequilibrium phase
transitions in the presence of external noise is still lacking.

In this Letter, we therefore study the effects of environmental
noise on two prototypical models of spreading and growth processes:
the space-independent (mean-field) logistic evolution equation and the spatially extended contact process.
In both cases, we find highly unusual behavior close to the extinction
transition.
It is characterized by a diverging effective noise
amplitude and enormous density fluctuations on long time scales,
motivating the name ``infinite-noise critical point''.
Infinite-noise critical behavior can be seen as counterpart of infinite-randomness critical behavior in
spatially disordered systems, but with exchanged roles of space and time.

In the remainder of this Letter, we first consider the logistic evolution equation
where a clear picture of the infinite-noise physics emerges from an asymptotically exact
random walk approach. To establish a framework for the analysis of temporal disorder,
we develop a real-time renormalization group (RG).
We then apply the RG to our second model, the contact process, and find that the DP critical
behavior of the  pure problem gets replaced by an infinite-noise critical point as well.
We confirm key findings by computer simulations, and we discuss applications.

\section{Logistic equation}

The logistic evolution equation
\begin{equation}
\dot \rho(t) =  [\lambda(t) -\mu(t)] \rho(t) - \lambda(t) \rho^2(t) ~.
\label{eq:mf-ODE}
\end{equation}
describes a variety of growth processes in nature, with applications
as diverse as biological population dynamics \cite{Verhulst38}, catalytic chemical reactions
\cite{SteinfeldFranciscoHase99}, and even linguistics
\cite{BodHayJannedy03}. In the context of
epidemic spreading, $\rho$ is the density of infected individuals,
$\mu$ is the rate at which sick individuals heal while $\lambda$ corresponds
to the rate at which a healthy individual is infected by sick ones.
If $\lambda$ and $\mu$ are time-independent, the behavior is easily understood;
the epidemic dies out exponentially for $\lambda<\mu$ and survives for infinite time
for $\lambda> \mu$. At the critical point, $\lambda=\mu$, the density
decays to zero, but only as a power law, $\rho \sim 1/t$.
Environmental noise, i.e., temporal disorder, can be introduced by making the healing and infection rates
time dependent. For definiteness, we consider the rates to be piecewise constant,
$\mu(t)=\mu_n$ and $\lambda(t)=\lambda_n$, over time intervals $\Delta t_n$.
The $\mu_n$ and $\lambda_n$ are independently
drawn from probability distributions $W_\mu(\mu)$ and $W_\lambda(\lambda)$.

\begin{figure}
\includegraphics[width=\columnwidth]{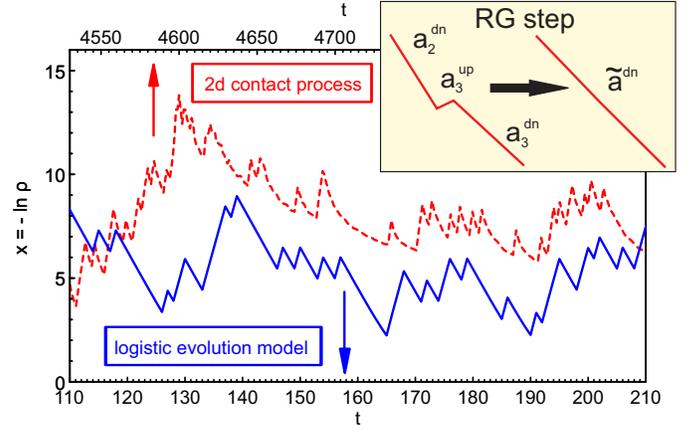}
\caption{(color online) Density $\rho(t)$ of individual noise realizations,
    plotted as $x=-\ln \rho$ vs.\ $t$.
    The logistic evolution results are obtained from eq.\ (1) with $\lambda=1, \Delta t=1$
    and binary disorder
    $W_\mu(\mu)=p\delta(\mu-\mu_h)+(1-p)\delta(\mu-\mu_l)$ with $\mu_h=2, \mu_l=0.5$ and $p=p_c=1/3$.
    The 2d contact process data stem from a Monte-Carlo simulation of  $1000^2$ sites
    using $\mu=1, \Delta t=2$ and
    $W_\lambda(\lambda)=p\delta(\lambda-\lambda_h)+(1-p)\delta(\lambda-\lambda_l)$ with $\lambda_h=3.65, \lambda_l=0.365$ and $p=0.8$.
    Inset: Schematic of the RG. In each step, the segment with the smallest change of $x=-\ln\rho$ is eliminated.  }
\label{fig:singleconf}
\end{figure}

If the healing and infection rates are time-independent, eq.\ (\ref{eq:mf-ODE}) can be solved in
closed form. Using this solution within each time interval of constant rates yields a linear
recurrence for the inverse density,
\begin{equation}
\rho_{n+1}^{-1} = a_n  \rho_{n}^{-1} + c_n~.
\label{eq:mf-recurrence}
\end{equation}
$\rho_n$ is the density at the start of time interval $n$.
The multipliers $a_n=\exp[(\mu_n-\lambda_n)\Delta t_n]$
reflect the exponential growth or decay
due to the linear term in
eq.\ (\ref{eq:mf-ODE}). The constants $c_n= (a_n-1)\lambda_n/(\mu_n-\lambda_n)$
are only important for large densities; they limit the growth and prevent $\rho_n>1$.
The time evolution is thus a random
sequence of decay and spreading segments during which the density $\rho$
either decreases or increases, depending on the balance between
$\mu(t)$ and $\lambda(t)$. This is illustrated in Fig.\ \ref{fig:singleconf}.
Neglecting the $c_n$ in the recurrence (\ref{eq:mf-recurrence}) for the moment,
we see that the logarithm of the density, $x_n= -\ln \rho_n$, performs a simple random walk,
$x_{n+1} =x_n +\ln a_n$.
The effect of the $c_n$ can be approximated by  a
reflecting boundary for this random walk at $x=0$. It limits the $x_n$ to positive values.

The reflected random-walk theory allows us to find the time evolution of the full probability
distribution of the density rather than just its average.
If the evolution starts from a fully infected system, $\rho_0=1$,
the probability distribution $P_n(x)$ after a large number $n$ of time intervals
is given by (for $x\ge 0$)
\begin{equation}
P_n(x)= \frac{2}{\sqrt{2\pi\sigma^2 n}} e^{-\frac{(x- v n)^2}{2\sigma^2 n}} -
\frac {2v}{\sigma^2} e^{ \frac {2 v x}{\sigma^2}} \Phi\left(\frac{-x-v n}{\sigma n^{1/2}} \right)
\label{eq:RW_P(x)}
\end{equation}
Here, $\Phi$ is the cumulative normal distribution, $v = \langle\ln a \rangle$ yields the drift of $P_n(x)$, and
$\sigma^2=\langle\ln^2 a \rangle -\langle\ln a \rangle^2$ gives its width.
[$\langle \ldots\rangle$ denotes the average over $W_\mu(\mu)$ and $W_\lambda(\lambda)$.]
The solution (\ref{eq:RW_P(x)}) can be verified by
inserting $P_n(x)$ into the drift-diffusion equation $\partial_n P = (\sigma^2/2)\partial_{x}^2P - v \partial_x P$
with flux-free boundary condition $v P - (\sigma^2/2)\partial_x P=0$ at $x=0$.

By inspecting the distribution $P_n(x)$, we can identify three different regimes:
$v<0$ corresponds to the active phase in which the epidemic survives for infinite time. $v>0$ is the inactive phase
in which the epidemic dies out, and $v=0$ is the critical point separating the two.
$v$ thus serves as a measure for the distance from criticality.
We now discuss the regimes in more detail.

In the active phase, $v<0$,  $P_n(x)$ approaches the stationary distribution
$P(x)=(2|v|/\sigma^2) \exp[-2|v|x/\sigma^2]$ in the long-time limit.
The distribution $P_\rho(\rho) = \rho^{-1+1/\kappa}/\kappa$ of the density
itself is highly singular and characterized by a non-universal
Griffiths exponent $\kappa=\sigma^2/(2|v|)$ that diverges at criticality.
The average stationary density is given by $\langle \rho \rangle = \langle \exp(-x) \rangle =1/(1+\kappa)$ while the
typical density reads $\rho^\mathrm{typ} = \exp(-\langle x \rangle) = \exp(-\kappa)$. Close to criticality,
the average density (which is dominated by rare events) is much larger than the typical one.

At the critical point, $v=0$, the distribution $P_n(x)$ simplifies to a
half-Gaussian, $P_n(x)= 2(2\pi\sigma^2 n)^{-1/2} \exp[-x^2/(2\sigma^2 n)]$, that
broadens without limit with increasing time. This clearly illustrates the notion
of infinite-noise criticality.
For long times, the average density thus decays as $\langle \rho_n \rangle =
\langle \exp(-x_n) \rangle = 2(2\pi\sigma^2 n)^{-1/2} \sim t^{-1/2}$.
In contrast, the typical density decays much faster, $\rho_n^\mathrm{typ} = \exp(-\langle x_n \rangle) = \exp[-(2 \sigma^2 n/\pi)^{1/2}]$,
implying  $\ln\rho^\mathrm{typ} \sim  -t^{1/2}$.

The inactive phase, $v > 0$, is more conventional as
the entire distribution $P_n(x)$ moves to larger $x$ with increasing time. Thus, the density
of almost all noise realizations rapidly vanishes such that both
average and typical densities
decay exponentially with time. Specifically $\rho_n^\mathrm{typ} \sim \exp(-v n)$ while $\langle \rho_n \rangle \sim \exp[-n v^2/(2\sigma^2)]$
for $v \le \sigma^2$ and $\langle \rho_n \rangle \sim \exp[-n(v-\sigma^2/2)]$  for $v \ge \sigma^2$.
The correlation
time $\xi_t$ is given by the time when the off-critical  $P_n(x)$
starts deviating significantly from its critical counterpart. This yields $\xi_t =(\sigma^2/v^2) \Delta t$.

We can cast these results in the language of critical phenomena
by comparing our results with the definitions of the critical exponents,
$\langle\rho\rangle \sim t^{-\delta}$ at criticality, $\langle \rho \rangle \sim |v|^\beta$
in the active phase, and $\xi_t \sim |v|^{-\nu_\parallel}$. The exponent values
\begin{equation}
\delta=1/2 ~,~\quad \beta =1 ~,\quad \nu_\parallel = 2~.
\label{eq:MF-exponents}
\end{equation}
differ from the ``clean'' ones
($\delta=1,~ \beta =1,~ \nu_\parallel = 1$ \cite{Hinrichsen00})
 but fulfill the scaling relation $\delta=\beta/\nu_\parallel$, and
$\nu_\parallel$ saturates Kinzel's bound $\nu_\parallel \ge 2$ \cite{Kinzel85}.  What about the correlation
length? The logistic equation does not contain any notion of space.
However, if it describes a $d$-dimensional system of
$N$ individuals, we can introduce a length scale $L\sim N^{1/d}$.
It is clear that the critical dynamics changes when the typical density
$\rho_n^\mathrm{typ} = \exp[-(2 \sigma^2 n/\pi)^{1/2}]$ becomes of order of $1/N$. This suggests an \
exponential dependence between correlation length $\xi$ and correlation time $\xi_t$,
\begin{equation}
\ln \xi  \sim \xi_t^{\omega} \qquad \textrm{with} \quad \omega=1/2~.
\label{eq:MF-dynamical scaling}
\end{equation}
The dynamical exponent $z$ is thus formally zero, and the correlation length exponent $\nu_\perp =\nu_\parallel / z$ is  infinite.
This highly unusual dynamical scaling is analogous to the activated scaling
at infinite-randomness critical points in spatially disordered systems \cite{Fisher92,Fisher95},
but with the roles played by space and time exchanged.

\begin{figure}
\includegraphics[width=\columnwidth]{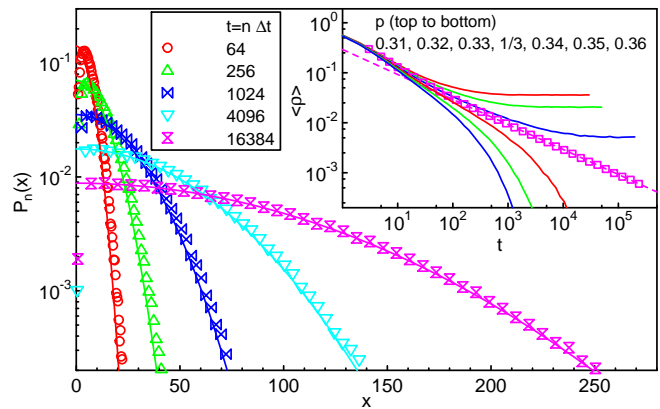}
\caption {(color online) Numerical solution of the logistic equation (\ref{eq:mf-ODE})
    for $\lambda=1, \Delta t=1$ using $10^6$ disorder realizations drawn from
    $W(\mu)=p\delta(\mu-\mu_h)+(1-p)\delta(\mu-\mu_l)$ with $\mu_h=2, \mu_l=0.5$.
    The transition is tuned by $p$, the critical point is at $p_c=1/3$.
    Main panel: Probability distribution $P_n(x)$ at criticality for
    several times $t$. The solid lines are the predictions of the random-walk theory, without adjustable
    parameters. Inset: Average density $\langle \rho \rangle$ vs.\ time $t$ for several
    $p$. The dashed line is a fit to the prediction $\langle \rho \rangle \sim t^{-1/2}$.}
\label{fig:mf}
\end{figure}

To illustrate and verify the reflected-random-walk theory, we have
solved the logistic equation (\ref{eq:mf-ODE}) numerically. Figure \ref{fig:mf}
shows the resulting $P_n(x)$ and $\langle \rho(t) \rangle$.
All data confirm our predictions. We have also verified
the stationary distribution $P(x)=(2|v|/\sigma^2) \exp[-2|v|x/\sigma^2]$
in the active phase close to criticality.

The knowledge of the full density distribution allows us to obtain many additional
results. The typical lifetime $\tau_N$ of a finite system of $N$ sites in the active phase
can be estimated from the probability that the density is below $1/N$. This yields
$\tau_N^{-1} \sim \int_0^{1/N} d\rho P_\rho(\rho) \sim N^{-1/\kappa}$. At criticality, the condition
$\rho^\mathrm{typ}(\tau_N)=1/N$ gives $\tau_N \sim \ln^2(N)$, and in the inactive phase
we find $\tau_N \sim \ln(N)$. All these lifetimes agree with the Langevin results of
Ref.\ \cite{VBLM11}.

\section{Real-time renormalization group}

To establish a general framework for treating the temporal disorder,
we now develop a real-time RG, in analogy to the strong-disorder RG
\cite{IgloiMonthus05} for spatially disordered systems.
We first combine consecutive time intervals with $\mu>\lambda$ into a single interval
 of length $\Delta t^\mathrm{up}$ (and consecutive intervals with $\mu<\lambda$ into a single interval of length
 $\Delta t^\mathrm{dn}$). This leaves us with a zig-zag curve of alternating
upward and downward segments for $x=-\ln\rho$, as illustrated in the numerical results of Fig.\ \ref{fig:singleconf}.
In each segment, $\rho^{-1}$ follows the linear recurrence
$\rho^{-1}(t+\Delta t) = a \rho^{-1}(t) + c$ with the multipliers of the upward segments fulfilling
$a^\mathrm{up}>1$ and those of the downward segments $a^\mathrm{dn}<1$.

The idea of the RG consists in iteratively eliminating the smallest upward and downward segments, coarse-graining time.
The distributions of $a^\mathrm{up}$ and $a^\mathrm{dn}$ change during this process, and their asymptotic
behavior governs the long-time physics.
In each RG step, we find the multiplier $a$
closest to unity. This defines the RG scale $\Omega = \min(a^\mathrm{up}_i,1/a^\mathrm{dn}_i)$.
We then replace this segment and its two neighbors by a single segment, as shown in the inset
of Fig.\ \ref{fig:singleconf}.
The time evolution of the
density in this renormalized segment follows from combining the three original density recurrences.
It takes the same linear form $\rho^{-1}(t+\Delta \tilde t) = \tilde a \rho^{-1}(t) + \tilde c$
but with renormalized coefficients $\tilde a$ and $\tilde c$. The renormalized multipliers
read
\begin{eqnarray}
\tilde a^\mathrm{up}&=&a_{i+1}^\mathrm{up} a_i^\mathrm{up} / \Omega~,
\label{eq:tilde_a_up}\\
1/ \tilde a^\mathrm{dn}&=&(1/a_{i}^\mathrm{dn})\, (1/a_{i-1}^\mathrm{dn}) / \Omega~,
\label{eq:tilde_a_dn}
\end{eqnarray}
for the decimation of downward segment $a_i^\mathrm{dn}$ and upward segment $a_i^\mathrm{up}$,
respectively.
The time intervals combine as
\begin{eqnarray}
\Delta\tilde t^\mathrm{up} = \Delta t^\mathrm{up}_{i} + \Delta t^\mathrm{dn}_i + \Delta t^\mathrm{up}_{i+1} ~,
\label{eq:tilde_t_up}
\\
\Delta\tilde t^\mathrm{dn} = \Delta t^\mathrm{dn}_{i-1} + \Delta t^\mathrm{up}_i + \Delta t^\mathrm{dn}_i ~.
\label{eq:tilde_t_dn}
\end{eqnarray}
The expressions from (\ref{eq:tilde_a_up}) to (\ref{eq:tilde_t_dn}) are exact. Moreover, they are equivalent
to the RG for the spatially disordered quantum Ising chain  \cite{Fisher92,Fisher95} and can be solved in the same way.

Following Fisher \cite{Fisher92,Fisher95}, we define logarithmic variables $\Gamma=\ln\Omega$,
$\beta=\ln a^\mathrm{up} -\Gamma$, and $\zeta=-\ln a^\mathrm{dn} -\Gamma$.
Upon iterating the RG steps,
$\Gamma$ increases, and the probability distributions of $\beta$ and $\zeta$,
 ${\cal R}(\beta;\Gamma)$ and ${\cal P}(\zeta;\Gamma)$ change.  They fulfill the integro-differential
(flow) equations
\begin{eqnarray}
\frac{\partial{\cal R}}{\partial\Gamma} & = & \frac{\partial{\cal R}}{\partial\beta}
                           +({\cal R}_0-{\cal P}_0){\cal R} +{\cal P}_0 \left({\cal R}\stackrel{\beta}{\otimes}{\cal R}\right) ~.
\label{eq:flow-R}\\
\frac{\partial{\cal P}}{\partial\Gamma} & = & \frac{\partial{\cal P}}{\partial\zeta}
                           +({\cal P}_0-{\cal R}_0){\cal P} +{\cal R}_0\left({\cal P}\stackrel{\zeta}{\otimes}{\cal P}\right)~.
\label{eq:flow-P}
\end{eqnarray}
where ${\cal R}_0={\cal R}(0;\Gamma)$ and ${\cal P}_0={\cal P}(0;\Gamma)$.
The symbol ${\mathcal{P}}\stackrel{\zeta}{\otimes}{\mathcal{P}}=\int_{0}^{\zeta}{\mathcal{P}}(\zeta^{\prime}){\mathcal{P}}(\zeta-\zeta^{\prime}){\mathrm{d}}\zeta^{\prime}$
denotes the convolution.

The complete solution of the flow equations for ${\cal R}(\beta;\Gamma)$ and ${\cal P}(\zeta;\Gamma)$
is quite elaborate \cite{Fisher92,Fisher95,Fisher94}. However, the physically relevant solutions can be obtained
from the ansatz
\begin{equation}
{\cal R}(\beta;\Gamma)={\cal R}_0 e^{-{\cal R}_0 \beta} ~, \quad
{\cal P}(\zeta;\Gamma)={\cal P}_0 e^{-{\cal P}_0 \zeta}~.
\label{eq:ansatz}
\end{equation}
Inserting this ansatz into the flow equations
(\ref{eq:flow-R}) and (\ref{eq:flow-P}) gives the ordinary differential equations
\begin{equation}
d{\cal R}_0 / d\Gamma =-{\cal R}_0 {\cal P}_0~, \quad d{\cal P}_0 / d\Gamma =-{\cal R}_0{\cal P}_0~
\label{eq:R0P0flow}
\end{equation}
which can be solved by elementary means.

This leads to three types of solutions for $\Gamma \to \infty$.
At criticality, $\langle \zeta \rangle = \langle \beta \rangle$
(equivalent to $v=\langle \ln a \rangle =0$), the distributions of the upward and
downward multipliers are identical, ${\cal P}(\zeta;\Gamma)=\exp(-\zeta/\Gamma)/\Gamma$ and ${\cal R}(\beta;\Gamma)=\exp(-\beta/\Gamma)/\Gamma$.
They broaden without limit with the RG scale $\Gamma$, i.e., the effective temporal disorder
diverges for long times, in agreement with the notion of ``infinite-noise" criticality.
The typical length of a renormalized time interval $\Delta \tilde t$ behaves as $\Gamma^2$.
In the inactive phase, $v = \langle \beta \rangle - \langle \zeta \rangle > 0$,
the distribution of the upward steps becomes very broad, ${\cal R}(\beta;\Gamma)={\cal R}_0 \exp(-{\cal R}_0 \beta)$ with ${\cal R}_0=\exp(-\Gamma/\kappa)/\kappa$
while the distribution of the downward steps does not broaden, ${\cal P}(\zeta;\Gamma)= \exp(-\zeta/\kappa)/\kappa$ with a constant $\kappa \sim 1/v$.
In the active phase, $v < 0$, the expressions for ${\cal P}(\zeta;\Gamma)$ and ${\cal R}(\beta;\Gamma)$ are exchanged. In both phases, $\Delta \tilde t$ scales as $\exp(\Gamma/\kappa)$.

Many important results follow from the RG. To estimate the lifetime of a finite-size
system of $N$ sites, we run the RG until the typical upward multiplier $a^\mathrm{up}=\exp(\Gamma+\beta)$ reaches $N$.
The lifetime $\tau_N$ is given by the typical time interval at that RG scale. Using the above solutions, we obtain
$\tau_N \sim N^{1/\kappa}$ in the active phase, $\tau_N \sim (\ln N)^2$ at criticality, and $\tau_N \sim \ln N$ in the inactive phase,
as in the random walk approach. Analogously, the width of the density distribution $P(x)$ at criticality is governed
by the upward multipliers $a^\mathrm{up}$. As the typical value of $\ln a^\mathrm{up}=\Gamma+\beta$ is $2\Gamma$, we obtain
a width of $\Delta x \sim \Gamma \sim t^{1/2}$ in agreement with (\ref{eq:RW_P(x)}) \footnote{The full $P(x)$
can be found from the joined distribution of $a$ and $c$ of the last segment at the end of the decimations.}.

\section{Contact process}

So far we have analyzed the logistic equation (\ref{eq:mf-ODE}) which contains density fluctuations
in time but not in space. What changes for finite-dimensional systems that fluctuate in space and time?
To answer this question, we turn to the contact process \cite{HarrisTE74}. It is defined on a $d$-dimensional
lattice. Each site can either be infected or healthy. During the time evolution, each infected site heals at
rate $\mu$ while healthy sites becomes infected at rate $\lambda n /(2d)$ where $n$ is the number of infected nearest
neighbor sites. (Note that the logistic equation (\ref{eq:mf-ODE}) can be understood as the mean-field
limit of the contact process.) If $\mu$ and $\lambda$ are uniform
in space and time-independent, the nonequilibrium transition between the active and inactive
phases is in the ubiquitous DP universality class \cite{GrassbergerdelaTorre79}.

Temporal disorder is again introduced by making the healing and infection rates random
functions of time; we assume piecewise constant rates,
$\mu(t)=\mu_n$ and $\lambda(t)=\lambda_n$, over time intervals $\Delta t_n$.
According to Kinzel's \cite{Kinzel85} generalization of the Harris criterion, such temporal
disorder destabilizes the DP critical behavior in all dimensions because the clean DP correlation
time exponent  $\nu_\parallel=\nu_\perp z$ violates the inequality $\nu_\parallel \ge 2$.
Temporal disorder is thus a relevant perturbation (having positive scale dimension) at the clean DP critical point.

To resolve the fate of the phase transition in the presence of temporal disorder,
we generalize our real-time RG method.
As in the logistic equation, the time evolution of the contact process is a sequence
of spreading and decay segments, depending on the  values of $\mu_n$ and $\lambda_n$.
If the temporal disorder is strong, the system is far away from criticality in each individual
segment. This allows us to neglect spatial fluctuations and to formulate the theory in terms
of the density $\rho(t)$ only. We will discuss the validity of this approximation in the conclusions.
During decay segments, the density decreases exponentially with time, as in the (mean-field)
logistic equation, because each site can heal independently. In contrast, the character of the
spreading segments changes because in a finite-$d$ system with short-range couplings, the infection
can at best spread ballistically,
$\rho(t) \approx \rho_0 (1+ b t)^d$, rather than exponentially as in the logistic equation.
This difference is clearly visible in Fig.\ \ref{fig:singleconf}.
(This also implies that $x=-\ln\rho$ does not undergo a simple random walk, consecutive
steps are rather correlated in a nontrivial manner.)

The RG must thus be modified. If a downward segment (density increase) is decimated,
the multipliers $a^\mathrm{up}$ still renormalize multiplicatively according to (\ref{eq:tilde_a_up}).
{If an upward segment is decimated, we need to combine the two neighboring downward (ballistic spreading)
segments during which the radii of active clusters grow linearly with
time. For strong disorder, $1/a_{i}^\mathrm{dn}, 1/a_{i-1}^\mathrm{dn} \gg \Omega$, we can therefore
estimate the renormalized multiplier as  $1/ \tilde a^\mathrm{dn}=(1+\tilde b\tilde \Delta t^{\rm dn})^d \approx (\tilde b\tilde \Delta t^{\rm dn})^d =
(b_i \Delta t_i^{\rm dn} + b_{i-1} \Delta t_{i-1}^{\rm dn})^d \approx
[(1/a_{i}^\mathrm{dn})^{1/d}+ (1/a_{i-1}^\mathrm{dn})^{1/d} ]^d$.
We thus arrive at an additive rather than a multiplicative renormalization
\begin{eqnarray}
(1/ \tilde a^\mathrm{dn})^{1/d}&=&(1/a_{i}^\mathrm{dn})^{1/d}+ (1/a_{i-1}^\mathrm{dn})^{1/d} - \Omega^{1/d}~
\label{eq:tilde_a_dn_fd}
\end{eqnarray}
Here, the last term contains the subleading contribution of
the decimated upward segment which we have mainly added to ensure the correct behavior in the
atypical case $1/a_{i}^\mathrm{dn} = 1/a_{i-1}^\mathrm{dn} = \Omega$
}
The RG defined by recursions (\ref{eq:tilde_a_up}) and (\ref{eq:tilde_a_dn_fd}) is equivalent
to that of spatially disordered quantum systems with super-Ohmic dissipation \cite{VHMN11} or with long-range
interactions \cite{JuhaszKovacsIgloi14}.
\footnote{{If the density increase in the spreading segments were slower than ballistic but still followed a power
in $t$, the recursion for $1/ \tilde a^\mathrm{dn}$ would take the additive form (\ref{eq:tilde_a_dn_fd})
with $1/d$ replaced by a different exponent. The resulting critical behavior would not change \cite{VHMN11}.}}

To analyze the RG, we define reduced variables $\beta=\ln a^\mathrm{up} -\Gamma$
and $\zeta=d[(\Omega a^\mathrm{dn})^{-1/d}-1]$. In terms of these variables, the flow equation for
the distribution ${\cal R}(\beta;\Gamma)$ is still given by (\ref{eq:flow-R})
while ${\cal P}(\zeta;\Gamma)$ fulfills
\begin{equation}
\frac{\partial{\cal P}}{\partial\Gamma}  =  \left(1+\frac {\zeta} d \right)\frac{\partial{\cal P}}{\partial\zeta}
                           +\left({\cal P}_0-{\cal R}_0+\frac 1 d \right){\cal P} +{\cal R}_0\left({\cal P}\stackrel{\zeta}{\otimes}{\cal P}\right)~.
\label{eq:flow-P_finite_d}
\end{equation}
In the mean-field limit $d\to \infty$, we recover the flow equation (\ref{eq:flow-P}), as expected.
Inserting the exponential ansatz (\ref{eq:ansatz}) into the flow equations (\ref{eq:flow-R}) and (\ref{eq:flow-P_finite_d})
yields
\begin{equation}
d{\cal R}_0 / d\Gamma =-{\cal R}_0 {\cal P}_0~, \quad d{\cal P}_0 / d\Gamma =(1/d-{\cal R}_0){\cal P}_0~.
\label{eq:R0P0flow_finite_d}
\end{equation}
These equations take the famous Kosterlitz-Thouless form \cite{KosterlitzThouless73}
for all finite $d$. Details of their solution will be published elsewhere. Here,
we just summarize the results.
The critical fixed point is located at ${\cal P}_0^\ast=0,{\cal R}_0^\ast=1/d$. This implies that
the distribution of the downward multipliers $1/a^\mathrm{dn}$ becomes infinitely broad while the
distribution upward multipliers $a^\mathrm{up}$ retains a finite width for $\Gamma \to \infty$.
The typical length of a renormalized time interval behaves as
$\Delta \tilde t \sim \Omega^{{\cal R}_0^\ast}=\Omega^{1/d}$.

Many important results follow.
The width of the (logarithmic) density distribution $P(x)$ at criticality is governed
by the typical value of $\ln a^\mathrm{up}$. This gives $\Delta x \sim \ln a^\mathrm{up} \approx \Gamma=\ln \Omega \sim \ln t$.
Thus, $P(x)$ still broadens without limit with increasing time, but its width increases only logarithmically, in contrast to the
power-law increase in the case of the logistic evolution equation.
The average density thus decays as $\langle \rho(t) \rangle \sim (\ln t)^{-\bar \delta}$ with $\bar\delta= 1$ (while the usual decay exponent $\delta$ formally vanishes).
The correlation time can be obtained from analyzing the off-critical solutions of (\ref{eq:R0P0flow_finite_d}). As usual for Kosterlitz-Thouless
flows, it depends exponentially on the distance from criticality $r$ via
$\ln \xi_t \sim  |r|^{-\bar\nu_\parallel}$ with $\bar\nu_\parallel=1/2$ (the usual correlation time exponent $\nu_\parallel$ is infinite,
fulfilling Kinzel's bound $\nu_\parallel \ge 2$).
Correlation length and time are proportional to each
other, implying a dynamical exponent $z=1$.
Finally, the stationary density in the active phase varies as $|r|^\beta$ with $\beta=1/2$. The critical exponents
\begin{equation}
\bar\delta=1 ~,~\quad \beta =1/2 ~,\quad \bar\nu_\parallel=1/2
\label{eq:finite-d-exponents}
\end{equation}
fulfill the scaling relation $\bar\delta=\beta/\bar\nu_\parallel$.
In the active phase, the lifetime $\tau_N$ of a
finite-size sample shows a Griffiths singularity, i.e., it increases as a nonuniversal power law $\tau_N \sim N^{1/\kappa}$.
However, in contrast to the logistic evolution equation, the Griffiths exponent $\kappa$ does not diverge at criticality but saturates at $\kappa_c=d$.

\begin{figure}
\includegraphics[width=\columnwidth]{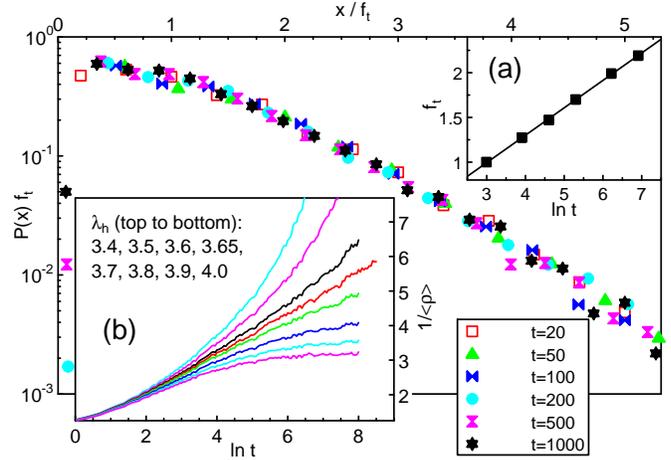}
\caption {(color online) Monte-Carlo simulations of the 2d contact process, starting from $\rho=1$
(up to $3200^2$ sites, $\mu=1, \Delta t=2$, $W(\lambda)=0.8\,\delta(\lambda-\lambda_h)+0.2\,\delta(\lambda-\lambda_h/10)$,
 20000 disorder realizations).
Main: $P(x)$ at criticality for different times $t$, scaled
such that the curves coincide. Inset (a): Scale factor $f_t$ vs.\ $\ln t$,
confirming the logarithmic broadening of $P(x)$ with time. Inset (b): $\langle\rho(t)\rangle$ for varying
infection rate $\lambda$. The critical curve, $\lambda_h=3.65$, follows the predicted $\langle \rho \rangle \sim (\ln t)^{-1}$ over almost
three orders of magnitude in $t$.}
\label{fig:cpt2d_all}
\end{figure}

To test these predictions, we have performed Monte-Carlo simulations of the 2d contact process with temporal disorder
by methods analogous to Ref.\ \cite{VojtaFarquharMast09}.
The time evolutions of $P(x)$ and  $\langle \rho(t) \rangle$ are shown in  Fig.\ \ref{fig:cpt2d_all}.
All data are in excellent agreement with the RG predictions. We have also verified
 $\tau_N \sim N^{1/2}$ at criticality.

\section{Conclusions}

In conclusion, random environmental noise (i.e., temporal disorder)
leads to exotic ``infinite-noise'' critical points at which the density distributions
become infinitely broad (even on a logarithmic scale) both for the (mean-field) logistic evolution model and
for the finite-$d$ contact process.
The infinite-noise critical behavior of the logistic equation is asymptotically exact because,
at criticality, the typical density vanishes with increasing time,
justifying the reflected-random-walk approach. Moreover, the real-time RG suggests
that this critical behavior holds
for an entire class of single-variable
growth models with temporal disorder in the linear growth rate and
some nonlinearity preventing population explosion. Such models occur in population dynamics,
chemical kinetics, economics, and other fields.

The generalization of the RG to finite dimensions includes spatial fluctuations only approximately (via the modification of the
density increase in spreading segments). The approximation is expected to be good if the temporal disorder
is so strong that individual spreading and decay segments are far from criticality.
As the density distribution broadens without limit at criticality, this condition
appears to be fulfilled self-consistently. Moreover, according to
Kinzel's \cite{Kinzel85} generalization of the Harris criterion, even weakly disordered systems
should flow to this regime.
{
These heuristic arguments suggest that our theory is asymptotically stable
for all bare disorder strengths. However, two other scenarios cannot be excluded: (i) our theory might be
asymptotically stable only for sufficiently strong bare disorder, or (ii) our theory might hold in a transient
time-regime (whose width diverges with disorder strength) before spatial fluctuations become important.
Discriminating between these scenarios requires more sophisticated methods.
}

Our theory for the contact process predicts universal critical behavior while Jensen \cite{Jensen96,Jensen05}
reported nonuniversal exponents in 1d. (Jensen's values do not always respect the bound  $\nu_\parallel \ge 2$, though.)
This could mean that our theory does not apply to 1d, or Jensen's data could be preasymptotic, caused by
long transients before the infinite-noise behavior is reached. We have performed Monte-Carlo simulations of the 1d contact process with
strong temporal disorder; preliminary results do show signatures of the strong-noise physics reported here.

In recent years, absorbing state phase transitions have been observed in turbulent liquid crystals \cite{TKCS07},
driven suspensions \cite{CCGP08,FFGP11}, superconducting vortex dynamics
\cite{OkumaTsugawaMotohashi11}, as well as in bacteria colony biofilms \cite{KorolevNelson11,KXNF11}.
Studying external noise at these transitions can provide experimental tests of our theory.
Moreover, the effects of noise on the extinction of a biological population or an entire species due to
environmental changes are attracting considerable attention (see, e.g., Ref.\ \cite{OvaskainenMeerson10}).
Experimentally, these questions could be
studied, e.g., by growing bacteria or yeast populations in fluctuating external conditions.

\acknowledgments

This work was supported by the NSF under Grant No.\ DMR-1205803,
 by Simons Foundation, by FAPESP under Grant No.\ 2013/09850-7, and by CNPq under Grant
Nos.\ 590093/2011-8 and 305261/2012-6. We acknowledge helpful discussions with R.\ Dickman.

\bibliographystyle{eplbibtv}
\bibliography{../../00Bibtex/rareregions}

\end{document}